**Review article**

# Multifaceted nature of defect tolerance in halide perovskites and emerging semiconductors

Irea Mosquera-Lois[1,†], Yi-Teng Huang[2,†], Hugh Lohan[2,1,†], Junzhi Ye[2], Aron Walsh[1,*] and Robert L. Z. Hoye[2,*]

[1.] Department of Materials, Imperial College London, Exhibition Road, London SW1 2AZ, United Kingdom

[2.] Inorganic Chemistry Laboratory, Department of Chemistry, University of Oxford, South Parks Road, Oxford OX1 3QR, United Kingdom

[†] These authors contributed equally

* e-mail: a.walsh@imperial.ac.uk (A.W.), robert.hoye@chem.ox.ac.uk (R. L. Z. H.)

## Abstract

Lead-halide perovskites (LHPs) have shot to prominence as efficient energy conversion materials that can be processed using cost-effective fabrication methods. A widely-quoted reason for their exceptional performance is their ability to tolerate defects, enabling long charge-carrier lifetimes despite high defect densities. Realizing defect tolerance in broader classes of materials would have a substantial impact on the semiconductor industry. Significant effort has been made over the past decade to unravel the underlying origins of defect tolerance to design stable alternatives to LHPs comprised of nontoxic elements. However, it has become clear that understanding defect tolerance in LHPs is far from straightforward. This review discusses the models proposed for defect tolerance in halide perovskites, evaluating the experimental and theoretical support for these models, as well as their limitations. We cover attempts to apply these models to identify materials beyond the lead-halide system that could also exhibit defect tolerance, and the successes and pitfalls encountered over the past decade. Finally, a discussion is made of some of the important missing pieces of information required for a deeper understanding and predictive models that enable the inverse design of defect tolerant semiconductors.

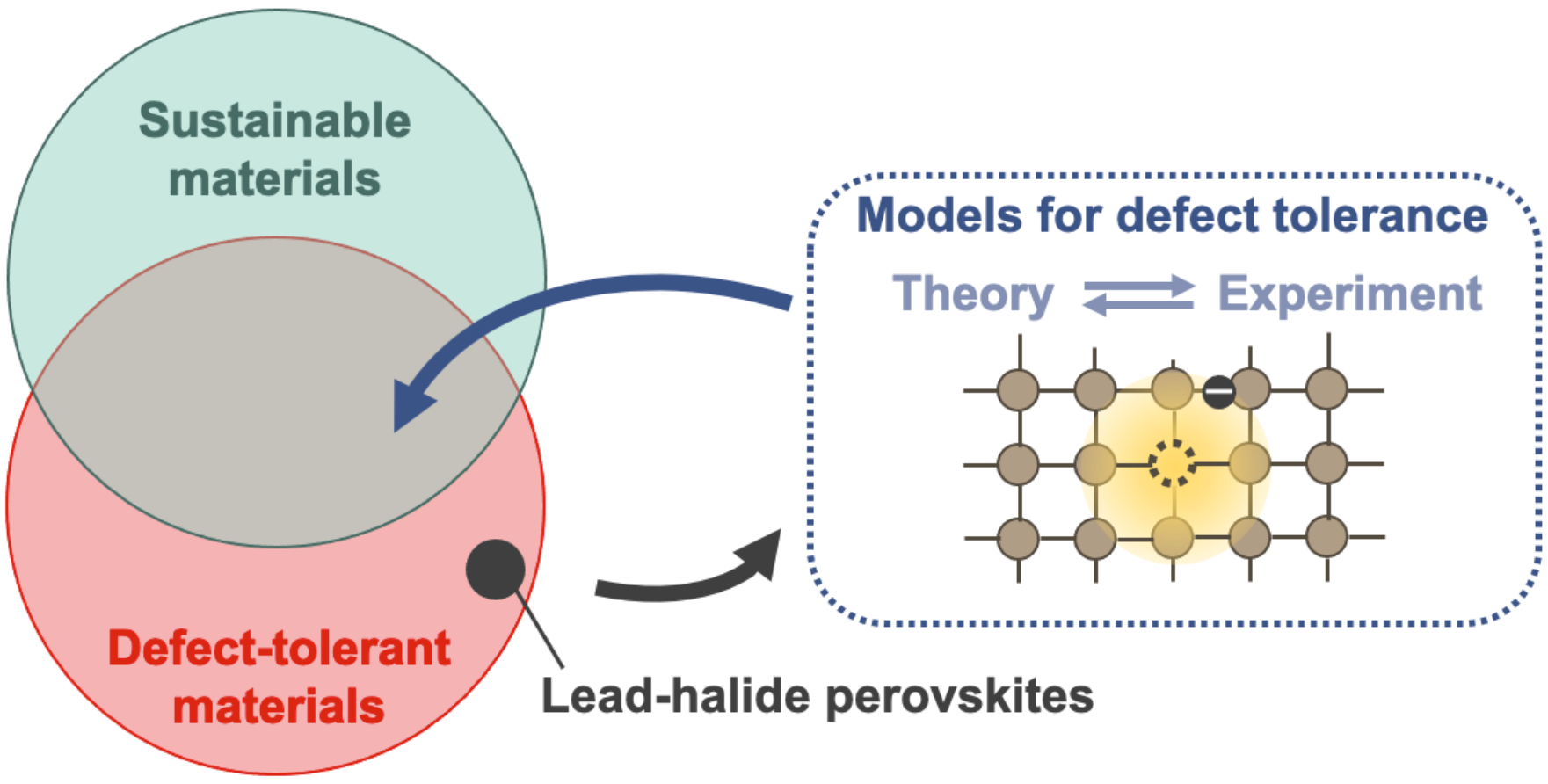

## Introduction

Semiconductors that harvest light to produce clean electricity[1,2], or clean fuels and chemicals[3–5], without emitting any greenhouse gases are becoming increasingly important for enabling society's transition to net-zero carbon dioxide equivalent ($CO_2$eq) emissions. The deployment of these technologies benefits from cost-effective manufacturing methods, but this typically leads to a compromise in performance because of the deleterious role played by defects[1]. Defects in semiconductors lower performance by causing irreversible losses in energy (refer to Box 1), as well as limiting the transport of charge-carriers. The effects of defects have traditionally been mitigated by minimizing their presence through careful materials growth and passivation[1,2,6,7]. But over the past decade, lead-halide perovskites (LHPs) have proven to be an exception, rapidly rising in performance in photovoltaics (PVs; Fig. 1a), which harvest light to produce clean electricity. The most efficient LHP PV devices are made using simple solution processing or evaporation methods, are polycrystalline, and the absorber layer is processed at temperatures up to only 150 °C[8,9]. By contrast, the most efficient Si PVs are made using capital-intensive equipment, are single crystalline, and processed at >1000 °C, and yet the certified record light-to-electricity power conversion efficiencies (PCEs) are very similar (27.3% for Si PVs, and 26.5% for LHP PVs at present)[8,9,10]. The ability to achieve efficient LHPs using standard laboratory equipment has engaged a large global community, such that the cycles of learning required to reach the performance of single-crystalline Si (c-Si) PVs has taken a significantly shorter period of time (Fig. 1a *vs*. b).

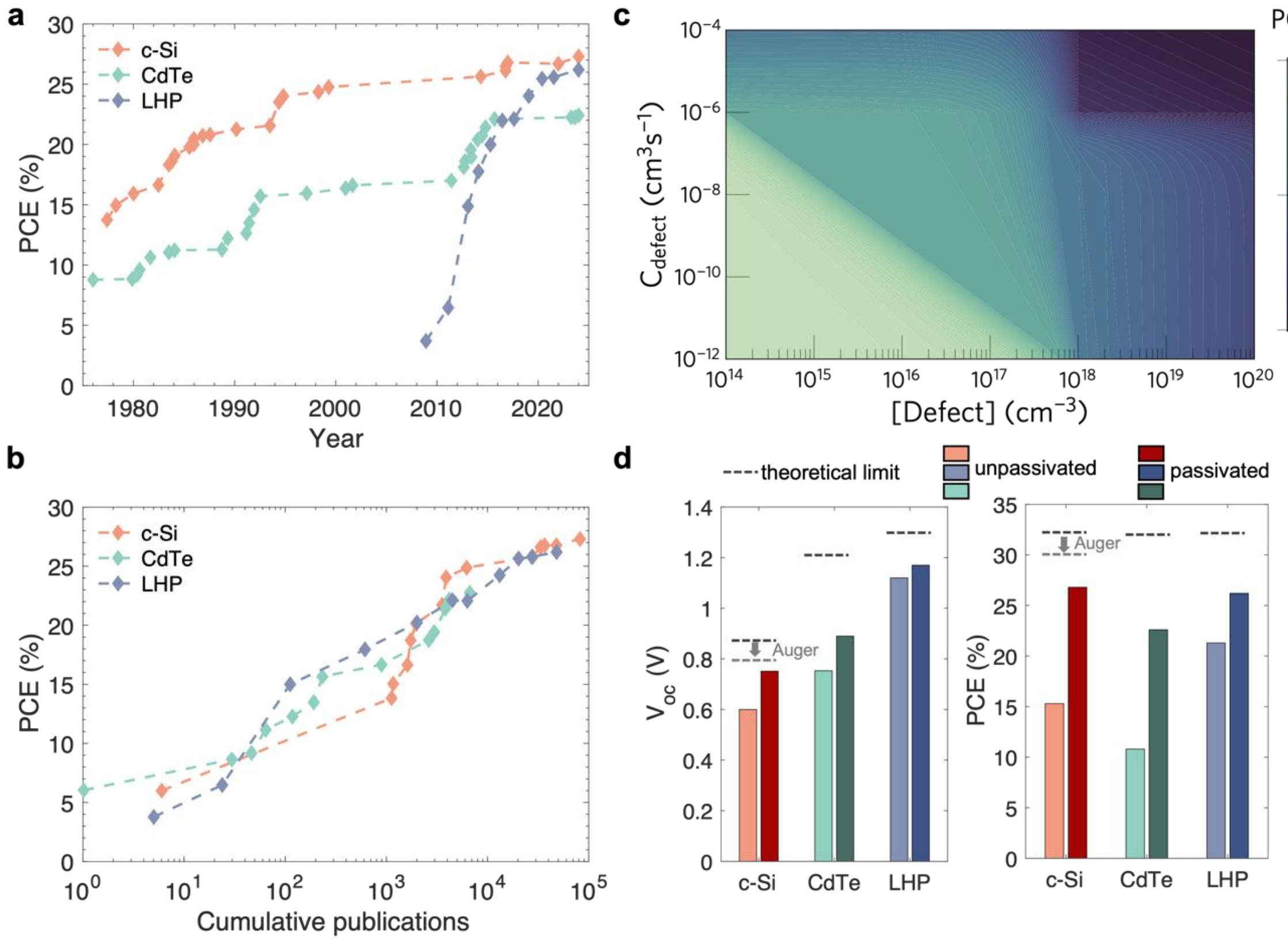

**Fig. 1 | Impact of defect tolerance on the performance of solar absorbers.** Comparison of the learning rate of lead-halide perovskite (LHP) *versus* crystalline silicon (c-Si) and cadmium telluride (CdTe) photovoltaics, with respect to **a,** development time and **b,** cumulative number of publications[9,11–14], which can be considered as a proxy for the level of effort. **c,** Map of the theoretical maximum power conversion efficiency of materials in PVs (PCE), depending on

the capture coefficient ($C_{defect}$) and concentration of defects. **d,** Loss in open-circuit voltage ($V_{OC}$) and power conversion efficiency of unpassivated and optimally passivated LHP, c-Si and CdTe photovoltaic devices compared to the radiative limit.

This unusual high-performance of LHPs has largely been attributed to defect tolerance, where high PCEs are achievable in solar cells despite a high defect density by primarily forming defects with low capture coefficients (Fig. 1c) [6,7,15–18]. Indeed, PVs based on polycrystalline LHP thin films can achieve 21.2% PCE without any dedicated efforts to intentionally passivate defects in the bulk of the perovskite layer or at interfaces[19], reaching 69% of the theoretical limit (Fig. 1d)[19,20]. By contrast, polycrystalline Si PVs could only reach ~17% PCE without significant interface passivation (Fig. 1d)[21–23]. If we compare LHPs with another polycrystalline direct-bandgap material, CdTe PVs could only reach ~10% PCE without passivation[24,14], while the optimally passivated devices are at only 70% of the theoretical limit—in contrast with 80% for passivated LHPs (Fig. 1d). Even in the most efficient LHP PVs, the defect density ($10^{13}$–$10^{17}$ cm$^{-3}$)[25–28] is orders of magnitude larger than in optimized single crystalline Si (<$10^{11}$ cm$^{-3}$)[29].

Defect tolerance in solar absorbers was discussed over 20 years ago by Zhang *et al.* to account for the surprisingly high efficiency of $CuInSe_2$ despite hosting extraordinarily high concentrations of native defects on the order of ~1%[30]. In this case, defect tolerance arises because the deep indium on copper antisites ($In_{Cu}^{2+}$) and copper vacancies ($V_{Cu}^{-}$) combine to form benign complexes ($2V_{Cu}^{-} + In_{Cu}^{2+}$) that can order to form new crystallographic phases[30]. But for LHPs, a wide range of models have been put forward to account for their tolerance to point defects, which draw upon electronic and structural factors[31–33]. More recently, the concept of defect tolerance has been extended by considering that the dynamic disorder of LHPs results in significant energetic fluctuations in trap levels, by as much as 1 eV on a picosecond timescale[34], raising the question of how traps can then be considered shallow or deep. Defect spectroscopy in LHPs is also complicated by contributions from ionic and electronic processes that are difficult to disentangle[35]. Deep-level transient spectroscopy measurements have indicated deep traps to be present[36,37]. Some of these defects are relatively benign with small capture cross-sections (~$10^{-15}$ cm$^2$)[36,37], while other works have identified more harmful defects with larger capture cross-sections ~$10^{-12}$ cm$^2$ [37]. On the other hand, using positron annihilation lifetime spectroscopy on methylammonium lead iodide ($MAPbI_3$), it was confirmed that positively charged defects (*e.g.*, $V_I^+$) have negligible trapping rates and are benign, whereas negatively charged lead vacancies in this perovskite were identified and found to act as recombination centres[38]. The concentration of these $V_{Pb}^{2-}$ was estimated to be high, in the range of $10^{15}$–$10^{17}$ cm$^{-3}$ [38]. Despite this, the non-radiative recombination coefficient reported for $MAPbI_3$ (1.4–1.5×$10^7$ s$^{-1}$) is similar to established inorganic semiconductors, such as c-Si (0.1–2.5×$10^7$ s$^{-1}$) and GaN (0.1–1.0×$10^7$ s$^{-1}$)[39].

The field has therefore reached a critical juncture, where it is important to bring together the past decade of experimental and computational work to evaluate the multifaceted nature of defect tolerance in LHPs. This review begins by discussing the definition of defect tolerance, which is inconsistent between communities because of the different ways in which defects are probed. Next, we evaluate the models put forward for defect tolerance, discussing the evidence for and against these models. We balance out the review with a discussion of efforts to generalize defect tolerance beyond LHPs, which can lead to the discovery of efficient, cost-effective solar absorbers that overcome the stability and toxicity limitations of halide perovskites. We discuss the strategies that have been adopted, as well as the successes and challenges found among the materials investigated. Finally, we discuss some of the important questions that need to be addressed to move forward towards a consolidated view of defect tolerance and the development of design rules that enable the discovery of defect tolerant semiconductors.

**Box 1 | Charge-carrier trapping at defect sites**

Point defects can create localized energy levels in crystalline materials. If these occur within the bandgap of a semiconductor, the associated charge transition levels (called traps) can capture electrons and/or holes, resulting in charge-carrier annihilation without emitting photons (*i.e.*, non-radiative recombination). Photoexcitation leads to the introduction of an excess carrier density with electrons ($e^-$) in the conduction band (CB) and holes ($h^+$) in the valence band (VB). The rate of charge trapping by defects will depend on this excess carrier density ($\Delta n$), along with both the concentration of active defects ($N_{\text{defect}}$) and their capture coefficients ($C_{\text{defect}}$), as shown in the left diagram below.

The microscopic process that determines $C_{\text{defect}}$ can be visualized with a configuration coordinate diagram (sketched below, right panel), where each curve represents the energy of the system ($E$) as atoms are distorted from their equilibrium configuration ($Q$)[39,40]. Starting with a neutral defect ($X^0$) in the presence of excess charge-carriers following photoexcitation (blue curve), by overcoming or tunnelling through the energy barrier $E_b{}^n$, an electron is captured, leading to a change in the defect charge state and distortion of the local structure (change in $Q$). The system becomes a negatively-charged defect with excess $h^+$ (red curve). Further overcoming energy barrier $E_b{}^p$ leads to $h^+$ being captured by $X^-$, and the system re-entering into the ground state (green curve). During this cycle, the energy of an electron and hole (*i.e.*, the bandgap $E_g$) has been converted into heat through multi-phonon emission.

The overall non-radiative recombination rate depends on the carrier with the lowest capture rate, as described by the Shockley-Read-Hall model[41]. It is possible to obtain a low rate despite a high defect density by having traps close to one of the band-edges, such that the capture rate of one charge-carrier is minimized (Fig. 1c). These are called shallow traps, and are one route to defect tolerance. While standard analysis for conventional semiconductors typically considers the regime of weak distortion (small changes in $Q$) and harmonic potential energy surfaces (*i.e.*, parabolic $E$–$Q$ curves), many defect charge transitions result in significant rearrangements in bonding (*e.g.*, dimer formation), especially for materials with soft structures. In this case, the energy gap between the trap level and band extrema is a poor proxy for charge-carrier capture, requiring consideration of the full potential energy surface[42].

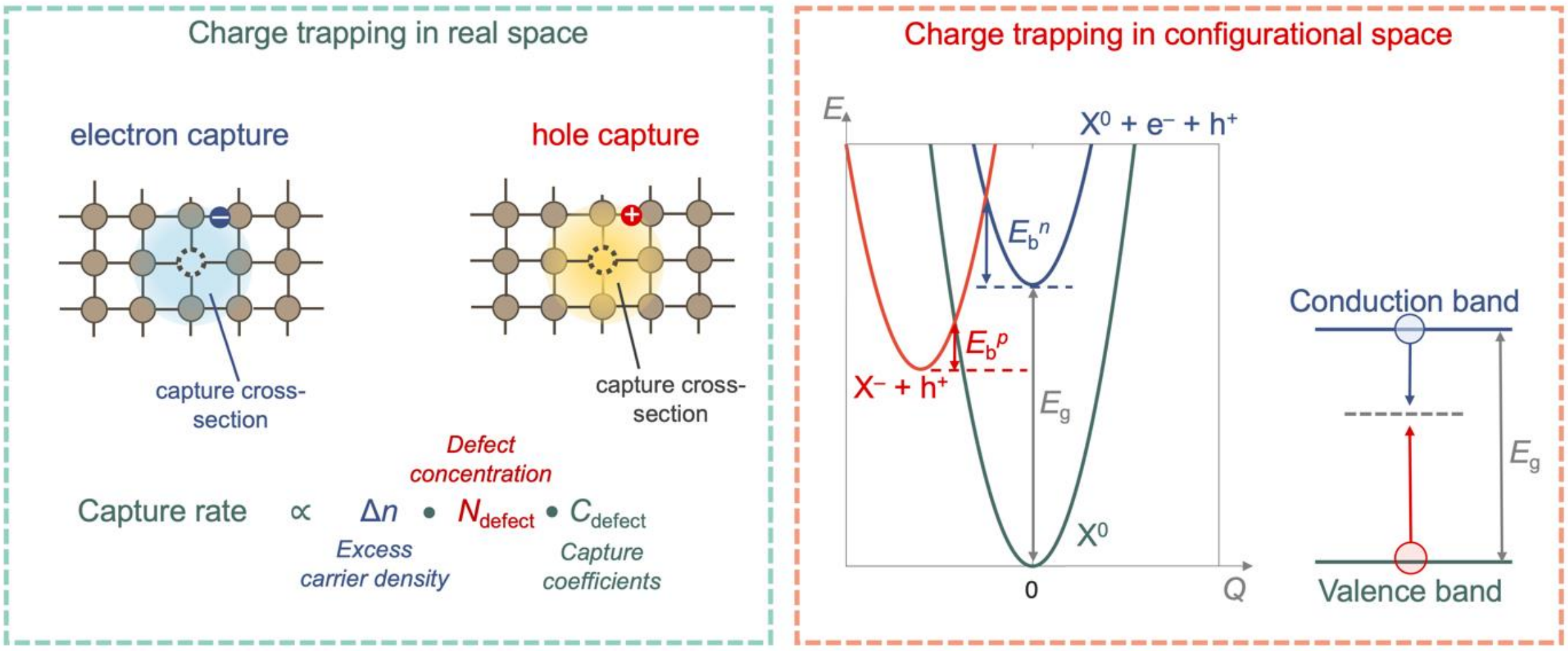

## Defect tolerance in halide perovskites

### *Defining defect tolerance*

A specific, widely-accepted definition of defect tolerance is currently elusive, partly because of continued debate over how defect tolerance arises, and partly because of the different probes used for evaluating the role of defects, from computational methods (Box 2) to experimental techniques, and with sensitivities ranging from the parts per trillion up to the percent level (Box 3). Adding to these difficulties, not all defects detected from measurements are recombination active. Analyzing defect concentrations alone without understanding what these species are makes it challenging to relate to non-radiative recombination rates.

Zakutayev *et al.* defined defect tolerance as "the tendency of a semiconductor to keep its properties despite the presence of crystallographic defects"[33]. Brandt *et al.* subsequently built upon this to specify that materials could be defect tolerant if they 1) form low defect concentrations despite being processed rapidly at low-temperature, or 2) have minimal reductions in charge-carrier mobility and minority carrier lifetime despite the presence of "extrinsic, intrinsic, or structural defects"[32]. The latter definition was to recognize the importance of long minority-carrier diffusion lengths to achieve efficient performance, and that charge-carrier transport has historically limited the development of emerging thin film solar absorbers[43]. It remains difficult to quantify "minimal reductions", but for the purposes of materials screening, Buonassisi and co-workers proposed to use 1 ns minority-carrier lifetime as a critical threshold, since established thin-film solar absorbers needed lifetimes exceeding this to eventually reach 10% PCE in PVs[44]. However, recent work has shown that overly relying on minority-carrier lifetime can be deceptive if self-trapped excitons or small polarons form, which can lead to prolonged decays in the measured photoexcited charge-carrier population, but give substantial reductions in mobilities, thus overall leading to short diffusion lengths [45,46,47]. We therefore propose to refine this definition to "the effect, where a semiconductor, which readily forms free carriers, does not experience a substantial increase in its non-radiative recombination rate or reduction in charge-carrier mobilities when defects are present in high concentrations", since materials with carrier localization, or which cannot readily separate excitons into free carriers, are inherently limited in performance.

*Classifying the role of defects*

A wide range of defect types exist in materials. These can be classified based on their scale and dimensionality. At the atomic scale, there are 0D point defects, comprised of one atom missing or misplaced (*e.g.*, vacancy, anti-site or interstitial). There are also structural defects, including 1D dislocations, 2D stacking faults and grain boundaries, and 3D twin domains, along with macroscopic voids, and defects occurring at surfaces and interfaces. Structural defects could be minimized through careful materials processing (*e.g.*, by increasing grain size or passivating surfaces), but 0D point defects are thermodynamically unavoidable, and are also the most straightforward to model (Box 2). The discussion of defect tolerance has therefore focussed on the effects of intrinsic point defects, and materials which tolerate point defects could also be more resilient to non-radiative recombination at structural defects.

The presence of a defect does not necessarily lead to an increase in non-radiative recombination. Defects must capture both an electron and hole to annihilate them. The role of defects in optoelectronic materials therefore strongly depends on the capture rates for charge-carriers, and this depends on the trap level and capture cross-section (Box 1), which describes the area around a defect where charge-carriers can be captured, typically in the range of $10^{-4}$–$10^{4}$ Å$^2$ [48,49]. Defects that do not introduce a trap level in the bandgap are recombination-inactive, and are therefore classified as benign defects (*e.g.*, $V_I^+$ in $CH_3NH_3PbI_3$)[50]. Such defects can give rise to defect tolerance; although, they may be active in other ways such as ion diffusion or chemical degradation. Similarly, defects with trap levels close to one of the band-edges and low capture-cross-sections are not recombination-active (Box 1), but will act as electron- (close to CBM) or hole-traps (close to VBM), with very low capture rates for the other charge-carrier. Recombination-active defects (termed recombination centres) tend to be defects with trap levels close to mid-gap, with similar capture coefficients for both charge-carriers, especially if the capture coefficients are high. Such 'killer defects', which can be detrimental even in low concentrations, include $V_{Se}$ in $Sb_2Se_3$[51] and $V_S$ in $Cu_2ZnSnS_4$[52].

Furthermore, defects can exist in multiple charge states, especially if the species involved have higher valence (*e.g.*, $V_{Bi}$ has more possible charge states than $V_{Ag}$), giving rise to multiple transition levels (Box 2). For example, while the (+/0) transition of $V_{Se}$ in $Sb_2Se_3$ is inactive (slow carrier capture), the (2+/+) transition is fast, thus rendering $V_{Se}$ a fast recombination centre[51]. Limitations of a single-trap model have been frequently observed, with the need to consider metastable charges or structural configurations that can also participate in the charge-trapping processes by introducing intermediate states[48,53,54].

**Box 2 | Predicting the role of point defects**

Defect behaviour can be modelled at the atomic scale. The addition, removal or rearrangement of atoms in a crystal can be described using quantum mechanical methods (*e.g.*, Density Functional Theory, DFT) or force fields (*e.g.*, classical or machine learning interatomic potentials). The key thermodynamic quantity that influences many derived properties is the defect formation energy ($\Delta H_f$).

The advantage of quantum mechanical methods is that the electronic structure is rigorously described and there is no *a priori* assumption on the type of bonding that occurs in the material or at the defect site. The main disadvantage is the high computational cost. Defect properties are also sensitive to choices in the level of theory, with a high level (*i.e.*, a hybrid exchange-correlation functional with relativistic effects) often required for accurate predictions of charge trapping with localized defect wavefunctions[28,39,55–58].

The first step of defect calculations consists of creating an atomic model for each native defect (*e.g.*, by removing a $Bi^{3+}$ ion to model $V_{Bi}^{3-}$ in $Bi_2S_3$). Since defects can capture charge-carriers from the host, thus undergoing structural rearrangements (see Box 1), a separate atomic model is needed for each charge state. Here it is key that each atomic model represents the most stable defect geometry for each charge state to obtain accurate predictions[53,59,60]. The next step involves calculating the formation energy of each defect, which describes how easily they can form. These energies can be combined to predict the equilibrium concentrations and thus the dominant defect[61,62]. There is progress in automating many of these steps in software packages[63–68] such as *doped*[69] and *pydefect*[70].

The charge transition levels, which correspond to states that defects introduce within the bandgap, can also be analyzed and are typically classified as resonant, shallow or deep. Generally, defects that introduce deep levels are considered to be potential recombination centres. However, the position of the defect levels is not a reliable proxy for recombination activity (Box 1)[39,52,71,72]. Instead, the non-radiative recombination rate for each defect can be calculated using the capture coefficients for electrons and holes (modelling the process depicted in Box 1)[39,49,73,74] with packages like *CarrierCapture*[75] and *NonRad*[76]. By considering the recombination rates for all detrimental defects, the overall effect on photovoltaic efficiency can be predicted[40].

Equilibrium growth conditions are commonly assumed, which can be used as an avenue to tune the defect populations (*e.g.*, growth temperature, ratio of the precursor elements or partial pressure). By predicting defect concentrations for varying synthesis conditions, one can identify the conditions minimizing the density of harmful defects[6,37]. Another factor that can affect the concentration of efficiency-killing defects is the incorporation of extrinsic dopants. For instance, impurities can reduce the concentration of detrimental defects by passivating them[77] (*e.g.*, by binding to the native defect to form a benign complex[30,78], like filling a vacancy site[51,79]). Alternatively, dopants can also act indirectly by modifying the density of free carriers (*i.e.*, the Fermi level), thus changing the concentration of the native defects[6] and their effect on non-radiative recombination[72,80].

### *Models for defect tolerance*

Several models have been proposed in the literature to explain one or more aspects of the observed defect tolerance of metal-halide perovskites.

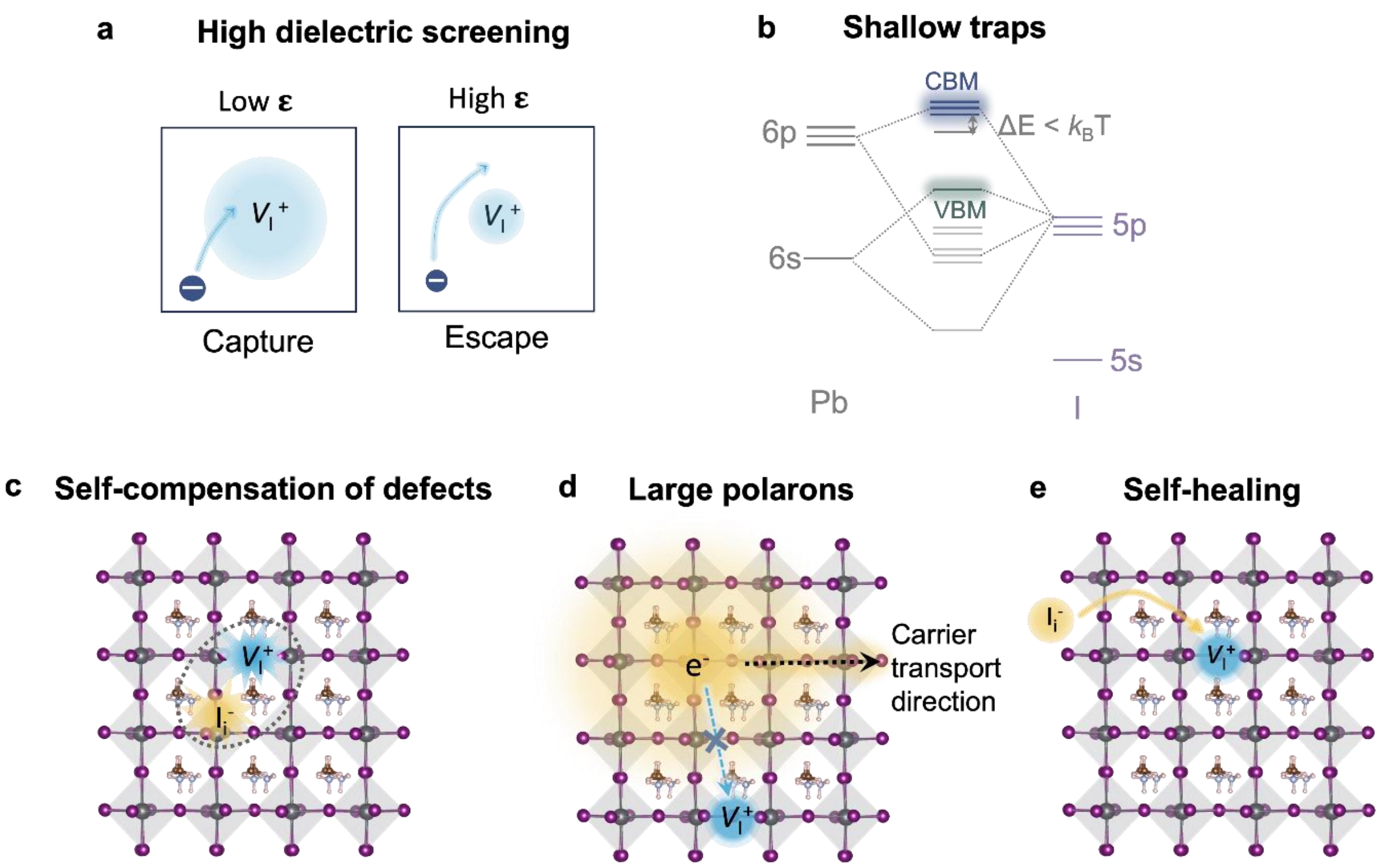

**Fig. 2 | Models for defect tolerance in lead-halide perovskites. a,** High dielectric screening, where the high dielectric constants reduce the interaction between free charge-carriers and

charged defects, thereby reducing the rate of carrier capture. **b**, Shallow trap model, where most traps are close to band-edges (within a few $k_BT$), and therefore benign. **c,** Self-compensation of defects, where defects form in charge-neutral combinations (*e.g.*, $V_I^+$ and $I_i^-$). **d,** Large polaron and low phonon energy model, where large polarons reduce the scattering and capture of charge-carriers by charged traps, and the low phonon energy leads to trapping being more difficult, with reduced nonadiabatic coupling constants. **e**, Self-healing, where fast ion migration promotes the annihilation of defects formed during exposure to damaging conditions.

**High dielectric screening.** The strength of the electrostatic interactions between charges in a crystal depends on the dielectric constant. This constant influences relevant processes, ranging from the binding energy between electrons and holes to form excitons, the scattering between charge-carriers and charged defects that may limit mobility, and the rates at which defects capture and annihilate charge-carriers. In standard inorganic semiconductors, the static dielectric constant is comprised of two components: a high-frequency optical response ($\varepsilon_{optic}$), and a low-frequency ionic response from phonons ($\varepsilon_{ion}$). For $CH_3NH_3PbI_3$, the reported values of $\varepsilon_{optic}$ (4–7) are typical for photovoltaic absorbers, while $\varepsilon_{ion}$ is substantially larger (17–29) due to the low-frequency infra-red active phonon modes. However, there is an additional component for the reorientation of the polar $CH_3NH_3^+$ molecule, which has been measured as between 13–37 depending on the temperature and frequency range[81]. Together the dielectric screening far exceeds the static dielectric constants of Si (12) or CdTe (10). These observations motivated the search for other highly polarizable semiconducting materials[32].

**Shallow trap model.** In 2014, Yin *et al.* reported defect calculations for $CH_3NH_3PbI_3$ that concluded: "dominant intrinsic defects create only shallow levels, which partially explain the long electron-hole diffusion length and high open-circuit voltage"[31]. The finding that the dominant point defects in halide perovskites do not introduce deep levels has held up well in subsequent studies[82,83]. This observation does not mean that deep states cannot form, for example, there is a significant body of literature on deep traps introduced by halide interstitials or lead antisites[71], but fortunately, they do not support rapid non-radiative recombination cycles[84,85].

Microscopically, the tendency to form shallow traps was attributed to the anti-bonding nature of the upper valence band (Pb 6s – I 5p) and lower conduction band (Pb 6p – I 5p), as illustrated in Fig. 2b. In this orbital arrangement, the atomic orbitals from which *non-bonding* defect states arise are located close to the band edges, and thus often lead to shallow or resonant states[6,86]. However, deep states can still form when there is significant hybridization between the defect dangling bonds, since the resulting (anti)bonding state will have (higher) lower energy than the original atomic orbitals[87]. Such behaviour is exemplified by the halide vacancy in the $CsPbX_3$ family, whose (+/0) donor level changes from shallow to deep through the halide series (I, Br, Cl)[88] due to the decreased lattice constant (thus smaller Pb–Pb distance in $CsPbCl_3$) and increased ionicity which favour the hybridization of Pb orbitals.[89,90] The relationship between increased lattice constant and reduced trap-mediated recombination has been further confirmed from carrier capture calculations of $I_i$ in orthorhombic FAPI, where a 1% lattice expansion reduces the carrier capture coefficient by one order of magnitude[91].

**Low density of deep defects.** An argument put forward against the shallow defect model is that some low-energy defects (*e.g.*, $V_H$ and $I_i$ in MAPI)[92,93] may act as rapid non-radiative recombination centres. However, these studies seem to be limited by either modelling less common defect species (*i.e.*, $V_H$ corresponds to forming methylamine, $CH_3NH_2$, which forms a gas that would evaporate under typical synthesis and processing conditions) or disagreements with other theoretical studies that also investigated carrier capture by $I_i$[84]. More importantly, they would challenge the experimentally observed long charge-carrier lifetimes found in LHPs. These discrepancies between theoretical studies illustrate the challenges for accurately modelling defects in LHPs due to the instability of the cubic phase at 0 K, and the soft potential energy surface that supports significant thermal motion, including dynamic octahedral tilting around room temperature.

**Self-compensation of defects.** Most LHPs are intrinsic semiconductors, with low charge-carrier concentrations in the dark[94], and are resistant to extrinsic *n*- or *p*-type doping. Sn-based compounds are an exception, due to the oxidation of Sn(II) to Sn(IV)[95]. One model used to explain the intrinsic behaviour is self-compensation. Rather than forming charged defects that are compensated by electrons or holes, the defects form in predominately charge-neutral combinations. Compensation can occur in the form of Schottky (vacancy) disorder, *i.e.*, $[V_A^-] + [V_B^{2-}] = 3[V_X^+]$, or Frenkel (vacancy/interstitial) disorder, *i.e.*, $[V_X^+] = [X_i^-]$[96]. The thermodynamic cost to form ensembles of compensated defects is low[96]. This introduces an insensitivity to the growth conditions and large defect concentrations can be supported without the detrimental effects of high carrier concentrations that would otherwise limit solar cell efficiency. This model has been further validated by posterior theoretical studies on MAPI, which predicted that the vacancies $V_I^+$, $V_{Pb}^{2-}$ and $V_{MA}^-$ form in high concentrations and follow a ratio ranging between 9.1:3.5:2 and 9.5:2.3:3.8 from MAI rich to $PbI_2$ rich conditions, respectively, thus in agreement with the formation of a stoichiometric amount of cation and anion vacancies[97]. In addition, the predicted high vacancy concentration supports the observed high ionic conductivities since they facilitate mass transport.

**Polaronic model.** Zhu and co-workers proposed that the defect tolerance of LHPs occurs due to the formation of large polarons[98–100]. Large polarons arise due to the weak to intermediate coupling between charge-carriers and longitudinal optical (LO) phonons, and there is now a plethora of evidence for large polaron formation from temperature-dependent mobility, optical pump terahertz probe spectroscopy (OPTP) and transient absorption spectroscopy and transient electron diffraction measurements, among others[47,98]. By forming large polarons, the long-range Coulombic potential experienced by charge-carriers from charged defects is weakened. Large polarons also have higher effective masses than free carriers, with the larger momentum reducing scattering by defect-induced phonons. Additionally, the shielding effect may add an extra energy barrier to the recombination of two large polarons with opposite charges, which slows the recombination process and increases the charge-carrier lifetime[1,8,9]. On the other hand, a disadvantage of forming large polarons is the reduction in charge-carrier mobility. Fortunately, the low deformation potential in electronically 3D halide perovskites ensures that small polarons do not form, such that charge-carriers remain delocalized, and mobilities > 50–200 $cm^2\ V^{-1}\ s^{-1}$ are achievable[101], which is important for realizing >1 μm diffusion lengths in polycrystalline thin films[102].

**Low-energy phonons.** A key feature of LHPs is the soft nature of their structure formed of flexible corner-sharing octahedra, with bulk moduli an order of magnitude smaller than metal oxides[103,104] (*e.g.*, 20±2 GPa for $MAPbI_3$ [105] *vs.* 390 GPa for $\alpha$-$Al_2O_3$ [106]). The underlying vibrational spectrum is complex, ranging from low-frequency modes (<100 $cm^{-1}$) associated with octahedral tilting and deformations to high-frequency modes (>100 $cm^{-1}$) linked to molecular vibrations[107]. The low-frequency infra-red active phonon modes make the largest contribution to electron-phonon coupling and result in an upper limit to the charge-carrier mobility[108]. Kirchartz *et al.* pointed out that within the Shockley-Read-Hall model of non-radiative recombination, a low phonon energy is beneficial. As carrier trapping is a multi-phonon emission process, a larger number of phonons need to be emitted if the phonon energy is smaller, making carrier trapping less likely to occur[109]. The role of phonon modes has also been explored using non-adiabatic molecular dynamics (NAMD) simulations that track the non-adiabatic coupling constant (NAC) between electrons/holes with defect states at finite temperatures. Low NAC values have been linked to defect tolerance including contributions from differences in wavefunction overlap, as well as contributions from the nuclear velocity of the heavy Pb and I atoms[103,104]. While NAMD is useful to model the dynamics of defects in excited states, it is important to consider all recombination mechanisms with appropriate charge-carrier and defect densities to estimate the contribution of changes in NAC to device performance. These factors can increase the predicted carrier lifetimes from ns to ms[110].

**Self-healing.** Self-healing refers to the ability of a material to autonomously recover its original performance after being damaged by an external stressor. It has been proposed to explain the high radiation resistance of LHPs[111–113], and their performance recovery during light-dark cycles[114–117]. Microscopically, self-healing has been attributed to rapid ion migration[26,111,114,118]. External perturbations, like high-energy radiation[111], light-illumination[114], stress[119], crystal cleavage[120], or chemical agents, can create defects. After the perturbation ceases, ions can migrate and return to their energetically favoured lattice positions, thus annihilating the defects[121]. Beyond defect migration, self-healing has also been linked to the low dissociation energy of LHPs into their binary constituents[122], which enables the rapid reformation of the LHP after damage. For example, the self-healing behaviour induced by chemical agents was investigated by Zhou and coworkers, who demonstrated that a mild stimulus (methylamine gas) can induce the formation of an intermediate liquid phase (*e.g.*, dissociation), which transforms into a smooth “defect-free” film after removing the gas[123].

**Box 3 | Measuring the point defects present and their effects**

The density, energy and capture cross-sections of defects, along with their effects on non-radiative recombination, can be determined using a wide range of spectroscopic, optical, electrical and capacitance techniques. Owing to the multi-varied nature of how each technique works, we focus on the overarching principles, the strengths, and challenges of using each method on the materials considered in this review.

Spectroscopic methods indirectly measure the effects of defects on non-radiative recombination, and have the advantages of being non-destructive and not requiring pinhole-free films. Common techniques include steady-state photoluminescence (PL) quantum yield (measuring the fraction of recombination events that are radiative)[124], and time-resolved techniques, such as time-resolved PL[125] and transient absorption spectroscopy[126]. With these methods, the sample is optically pumped, but other pump sources can be used, *e.g.*, electrons in cathodoluminescence spectroscopy[127], and positrons in positron annihilation spectroscopy[38]. Often these techniques are used together, but extracting recombination rate constants requires the development of a model to fit the time-resolved data[44,128], which is typically a simplification of the many complex processes occurring.

Relative changes in defect concentrations can be determined by measuring energetic transitions below the bandgap. However, the increase in sub-gap absorption due to defects is typically orders of magnitude below band-edge absorption. Techniques sensitive to these small changes include photothermal deflection spectroscopy[129], surface photovoltage measurements[130], and Fourier transform photocurrent spectroscopy[131]. However, these techniques are unable to directly quantify defect densities or identify the defects present, and are more useful for qualitative comparisons between samples.

Quantification of defect concentrations can be achieved through space-charge limited current density measurements[132]. This method requires a single-carrier device to be made, and the voltage swept until traps are filled. The trap density can be determined from the voltage at which trap filling occurs, but this assumes an ideal device, which often does not occur. More accurate ways to quantify defect density, as well as capture cross-sections, involve monitoring changes in the capacitance of a device due to trap filling/de-trapping. Prominent techniques include deep-level transient spectroscopy (DLTS)[133] and admittance spectroscopy[134]. For example, DLTS is sensitive to defects at the parts per trillion level[135]. However, these techniques cannot directly identify the defects giving rise to traps, require the fabrication of devices (*i.e.*, requires compact films), and are unable to detect shallow defects[133,134]. Ion migration will affect the capacitance profiles, requiring consideration of how these affect the measurements[35]. Furthermore, DLTS/admittance spectroscopy may present lower estimates of trap densities, and need to be treated with caution.

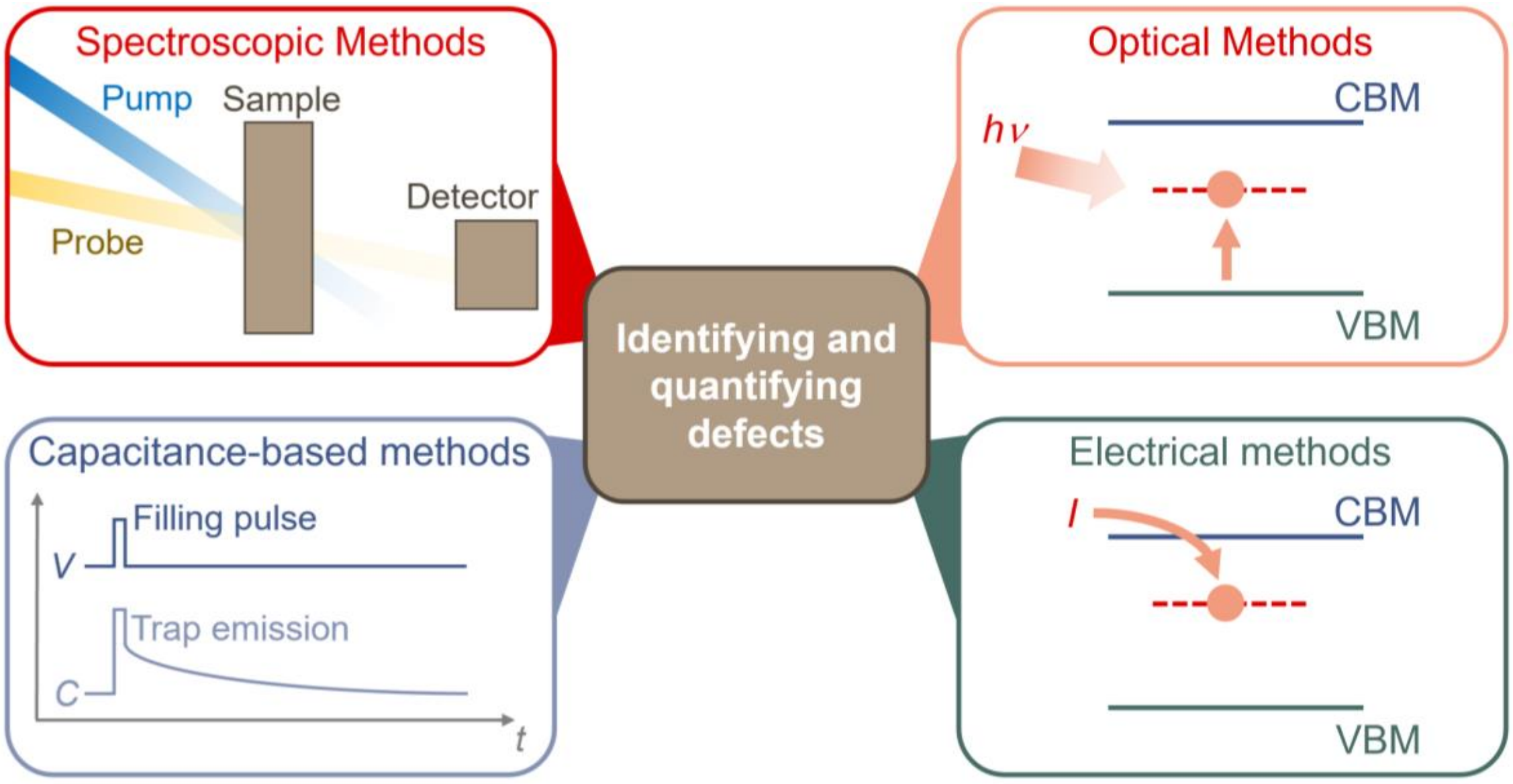

## Challenges in generalizing defect tolerance

A core motivation for unravelling the origins of defect tolerance in LHPs is to replicate this feature in broader classes of semiconductors, particularly nontoxic and stable compounds. Semiconductors developed on this basis are termed ‘perovskite-inspired’, and have especially focussed on materials with heavy main group cations that have stable valence $ns^2$ electron pairs ($In^{+}$, $Sn^{2+}$, $Sb^{3+}$, $Bi^{3+}$). In this section, we cover key points in the progress of these materials in PVs, before discussing the challenges in obtaining defect tolerance in chemistries beyond halide perovskites.

### *Progress of Perovskite-Inspired Materials (PIMs) in Photovoltaics*

To give a sense for the progress in this area, we cover key perovskite-inspired materials (PIMs), namely Sn and Ge perovskites, Bi-halide compounds, and antimony chalcogenides. These are most relevant for the discussion below. More detailed coverage of this broad materials space, including In-based compounds, can be found in Refs. 136–140. Whilst the exploration of some compounds predates the development of LHP PVs, it is useful to consider these as PIMs because they are electronically, structurally or chemically analogous to LHPs, and the motivation is the same – to develop an efficient, nontoxic and stable alternative to halide perovskite PVs.

**Sn and Ge perovskites.** The greatest success has been in materials most similar to LHPs, the isostructural tin-based perovskites ($ASnI_3$, $A \in \{MA, FA, Cs\}$). $FASnI_3$ PVs have now reached PCEs of 15.7% (Fig. 3), with a PL lifetime of 207 ns. However, Sn perovskites have limited stability, due to the facile oxidation of Sn(II) to Sn(IV), forming reactive defects that accelerate degradation. As such, excess $SnX_2$ halides and reducing additives, such as Sn powder or hydrazine vapour, and its derivatives are widely used in high-efficiency Sn perovskite solar cells[141]. Careful encapsulation of Sn perovskite devices, or the formation of protective layers (*e.g.*, $SnCl_2$ covering grains) are essential[142]. Similarly, Ge(II) is easily oxidized, leading to the formation of detrimental defects and competitive phases[143,144], and with few exceptions (*e.g.*, $CsGeI_3$) have bandgaps exceeding 2 eV[145]. There is therefore limited work on Ge-based perovskites.

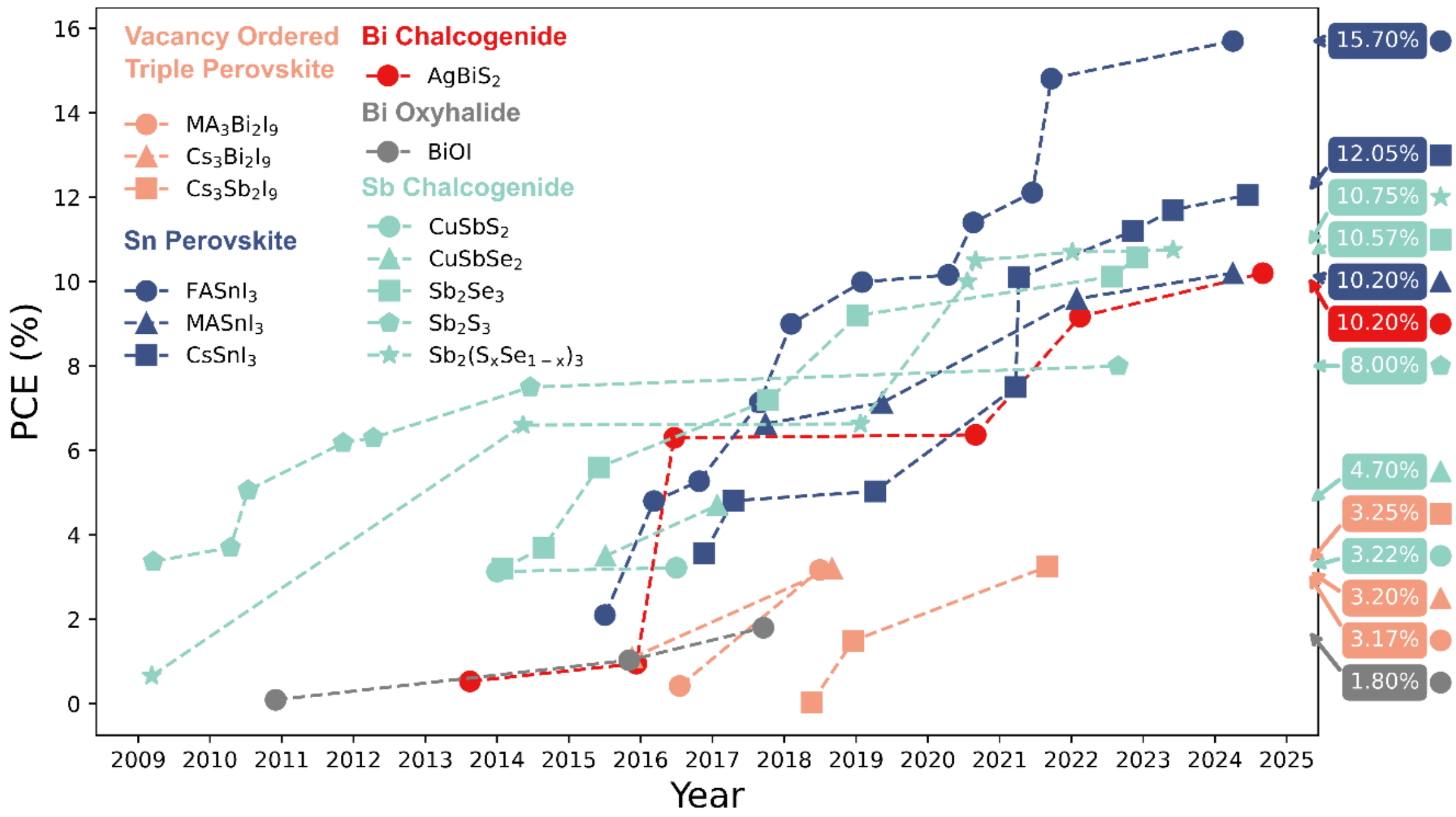

**Fig 3 | Progress of perovskite-inspired materials as solar absorbers**. Plot of the progress in the photovoltaic power conversion efficiency (PCE) over time of key compounds. References by material: $MA_3Bi_2I_9$[146,147], $Cs_3Bi_2I_9$[148,149], $Cs_3Sb_2I_9$[150–152], $FASnI_3$[153–163], $MASnI_3$[164–167], $CsSnI_3$[168–175], $AgBiS_2$[176–181], BiOI[182–184], $CuSbS_2$[185,186], $CuSbSe_2$[187,188], $Sb_2S_3$[189–195], $Sb_2Se_3$[196–201], $Sb_2(S_xSe_{1-x})_3$[202–208]. Refer to Supplementary Table 1 for details.

**Bi-halide materials.** While Bi-based materials have been the subject of much research due to their low toxicity[209], a common characteristic is self-trapping, which substantially limits mobilities and diffusion lengths, gives rise to larger open-circuit voltage losses, and can lead to unavoidable energy loss channels (see carrier-phonon coupling section for further discussion). As such, there is poor performance among many Bi-based materials, *e.g.* $MA_3Bi_2I_9$ [147], $NaBiS_2$ [210], and $BiI_3$ [211], all of which form small polarons or self-trapped excitons. However, recent work has shown self-trapping is not universal among Bi-based materials, and both BiOI and homogenously-disordered $AgBiS_2$ have mobilities significantly above unity[212,213]. There is therefore hope that Bi-based PIMs could be developed into efficient devices, and indeed $AgBiS_2$ has already reached a certified PCE of 8.85%[177], with lab-measured PCEs now exceeding 10% (Fig. 3)[181].

**Antimony chalcogenides.** The isostructural antimony chalcogenides, $Sb_2Se_3$, $Sb_2S_3$ and $Sb_2(Se,S)_3$, are promising systems, with highest reported efficiencies of 10.57%[214], 8.00%[189] and 10.75%[208], respectively, but are limited by low $V_{OC}$. Of the binary compounds, the selenide analogue shows the best efficiencies, where process control of the defects present is key to high-efficiency devices. There has been some debate over whether the critical performance losses originate from deep defects or self-trapping (see carrier-phonon coupling section). Nevertheless, reducing the grain boundary density, and passivating interfaces (*e.g.*, through the use of chelating or lanthanide additives) has been beneficial for improving $V_{OC}$ and device performance[215,216].

Ternary cuprous compounds include $CuSbSe_2$ and $CuSbS_2$, both of which showed early promise in their performance, but have since stagnated (with peak PCEs of 4.7%[186] for $CuSbSe_2$ and 3.2%[187] for $CuSbS_2$; Fig. 3). These isostructural materials have been predicted to be defect tolerant[217]. For both materials, a Cu-poor growth condition is preferred, as the dominant $V_{Cu}$ results in p-type character *via* self-compensation. Poor performance has been attributed to low $J_{sc}$ values, yet the reasons remain unexplored[218]. Morphological defects, such as voids and delamination, are also common and device-limiting. Additionally, it has been proposed that grain boundary defects may be strongly detrimental in this class, as with CdTe[24,14].

***Challenges in Achieving Defect Tolerance in Perovskite-Inspired Materials***

**Diversity of the electronic structure.** Early efforts to discover PIMs were driven by the dielectric screening and shallow trap models (Fig. 2a,b)[32,44]. However, these models have not been able to pinpoint the discovery of defect tolerant semiconductors. For example, Zakutayev *et al.* proposed an electronic structure model for defect tolerance in $Cu_3N$ resembling the model for LHPs. In $Cu_3N$, cation-anion orbital interactions produce a pair of bonding-antibonding states in both the upper valence and lower conduction bands (Fig. 4a)[33]. Although this was proposed to increase the likelihood of forming shallow defects[33], recent work has suggested $Cu_i$ as a potential deep trap[219]. LHPs similarly form a bonding-antibonding pair within the upper VB, although the CBM has an antibonding state (Fig. 2b). Such an electronic structure improves the likelihood of shallow acceptor defects. Buonassisi and co-workers therefore proposed to search for PIMs where the upper valence band has a significant contribution from the cation valence $s^2$ electrons, as they could result in similar bonding-antibonding states as in LHPs[32,44].

However, this qualitative model does not account for the degree of interaction between cation/anion orbitals or the specific crystal environments. In particular, Bi has a higher effective nuclear charge than Pb, and this results in the $6s^2$ orbital being deeper and more misaligned with the anion orbitals, resulting in a less disperse VB (higher effective mass), along with an increased tendency to form deeper traps.

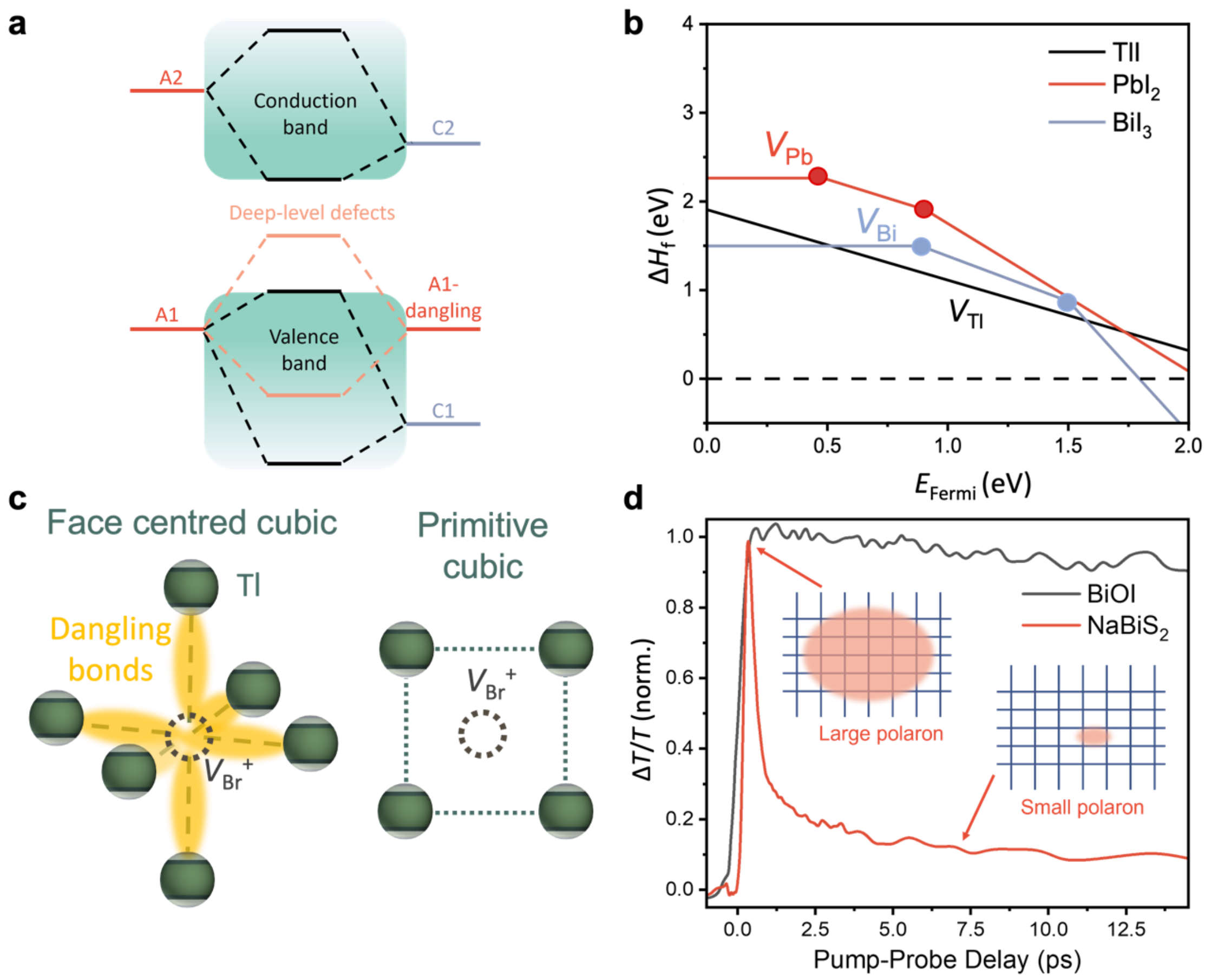

**Fig 4 | Performance bottlenecks in emerging perovskite-inspired materials**. **a,** Energy misalignment leading to the presence of deep defects. A, A-dangling, and C refer to the anion, anion dangling bond, and cation orbitals, respectively. **b,** Calculated formation energies D$H_f$ of the cation vacancies in TlI, $PbI_2$, and $BiI_3$[220]. **c,** Strong overlap of dangling bonds due to $V_{Br}^+$ in a face centred cubic lattice of TlBr (left), *vs.* weak dangling bond overlap in the primitive cubic polymorph (right). **d,** Normalized photoconductivity kinetics of BiOI and $NaBiS_2$ [210,213]. In $NaBiS_2$, delocalized large polarons initially occur after photoexcitation, but due to strong electron-phonon coupling, small polarons rapidly form on a ps timescale, leading to localized small polarons and a sharp decrease in photoconductivity. By contrast, the photoconductivity of BiOI slowly decays because carrier localization does not occur.

**Strength of interaction between dangling bonds**. When defects form, dangling bonds can hybridize, forming bonding-antibonding states. The (anti)bonding states can occur within the bandgap, becoming recombination-active traps (Fig. 4a). Stronger hybridization leads to greater bonding-antibonding state splitting, increasing the likelihood of deep traps. This hybridization strength between dangling bonds depends on their spatial and energetic separation. Therefore, traps become deeper when the lattice parameter decreases (see shallow trap model sub-section), or when the cation-anion orbitals are more poorly aligned (*i.e.*, as C1 is further from A1 in Fig. 4a). For example, in comparing iodides of heavy main group cations with stable $6s^2$ electron pairs (Tl to Bi), the cation vacancy forms a deeper level going across this series (Fig. 4b) due to the increasing energetic separation between cation-anion orbitals (C1-A1 in Fig. 4a). Furthermore, the higher cation oxidation state leads to a larger number of transition levels (Fig. 4b), thus increasing the likelihood of deep traps[220]. The role of spatial separation can be illustrated by BiOI, where the long Bi-I bonds (3.4 Å) weaken the interaction between Bi dangling orbitals, thus forming a shallow $V_I$, in contrast with the deeper $V_O$ and the shorter Bi-O bonds (2.3 Å)[184,221].

The strength of interaction between dangling bonds is also affected by the ionic radius and anion coordination number. For instance, large anions coordinated with small cations, as well as low anion coordination numbers, have larger inter-cation separation, which could result in shallow anion vacancies[220].

The discussion thus far has been on compounds where the cation/anion orbitals hybridize. Another way to form shallow traps involves materials with little interaction between the cation and anion orbitals, such that forming vacancies would not lead to dangling bonds. This could be obtained either by forming a structure with separate cation/anion sub-lattices (*e.g.*, TlBr in the primitive cubic rather than face centred cubic lattice; Fig. 4c)[222], or by having cation/anion orbitals with sufficient energetic separation to avoid mixing (*e.g.*, monolayer transition metal dichalcogenides)[223].

**Low dimensionality**. A critical difference between most of the PIMs explored and LHPs is the lower structural and electronic dimensionality of PIMs, leading to important challenges. Firstly, excitons can form more easily, impeding charge-carrier extraction. Secondly, lower dimensionality favours wider bandgaps, with values close to 2 eV[224] for most PIMs *vs.* 1.5–1.6 eV for $MAPbI_3$[225,226]. A wider bandgap reduces the optical dielectric constant since fewer free carriers are available for polarization, thus reducing the dielectric screening of charged defects. Thirdly, in some cases, low dimensionality could facilitate defect formation. For example, although $Sb_2Se_3$ has been considered promising because it has achieved >10% PCE in PV devices (Fig. 3), both DFT calculations and DLTS measurements indicate its lack of defect tolerance[51,227,228]. This has been attributed to the quasi-1D crystal structure of $Sb_2Se_3$, where the $[Sb_4Se_6]_n$ chains are bound through *van der Waals* interactions. These weak interactions, in addition to the valence alternation of Sb and Se[229], allow significant structural reconstructions upon defect formation, thus stabilizing many charge states and leading to many defect levels. Finally, low electronic dimensionality also increases the likelihood of carrier localization, as will be discussed next.

**Carrier-phonon coupling.** Charge-carriers can couple with lattice vibrations, giving rise to quasi-particles known as polarons[230]. Strong electron-phonon coupling can localize the carrier wavefunction within a unit cell, which is known as carrier localization or self-trapping. This substantially reduces mobilities (typically to <1–10 $cm^2\ V^{-1}\ s^{-1}$), limiting diffusion lengths, and is unavoidable even in defect-free materials. The past decade of work has revealed the significant presence of self-trapping in PIMs, especially Bi-based materials[46,231,232], including $BiI_3$ [233], $NaBiS_2$ [210], Cu-Ag-Bi-I compounds[211,234], $Cs_2AgBiBr_6$ [231], and other silver-bismuth elpasolites. This originates from their low electronic dimensionality, which can lead to barrierless carrier localization, especially when there is strong carrier-acoustic phonon coupling. Such a process can be observed experimentally by monitoring the photoconductivity, where the signal significantly decreases within a ps, as demonstrated in the case of $NaBiS_2$ (Fig. 4d).

Carrier localization can change the optoelectronic behaviour of materials, such that deep trap states no longer play a dominant role. For example, introducing defects at the percentage level in $NaBiS_2$ barely changed the decay of the excited state, which proceeded on the microsecond timescale[210]. This is due to the formation of small electron and hole polarons, spatially separated on Bi and S sites, respectively, which lowers the likelihood of both charge-carriers annihilating *via* the same defect. Whilst this gives the appearance of defect tolerance, strong localization leads to low charge-carrier mobilities (<0.1 $cm^2$ $V^{-1}$ $s^{-1}$; Fig. 4d), thus limiting diffusion lengths and charge-carrier extraction. Developing efficient PIMs therefore necessitates compounds with free charge-carriers. By studying $CuSbSe_2$, we rationalized that this may be found in compounds with a layered structure. This results in a lower deformation potential, because the strains to the unit cell from a propagating acoustic wave are mostly relaxed through changes in inter-layer gaps, rather than bond lengths. Furthermore, orbital hybridization across inter-layer gaps results in a higher electronic dimensionality (*e.g.*, $CuSbSe_2$ has a nearly 3D lower CB), reducing the likelihood of carrier localization being energetically favourable.

Finally, electron-phonon coupling can also create a non-radiative loss channel that does not arise from defects. We found this in BiOI, where the coupling between charge-carriers and interlayer breathing modes significantly distorts the potential energy surface, such that the ground and excited states approach each other, and excitations can then directly return to the ground state following a non-radiative pathway.

## Conclusions and outlook

Defect tolerance is key to developing high-performance and cost-effective optoelectronic materials. The exceptional performance of LHPs has revived this topic, which has been examined from a wide range of perspectives. Although this has led to different interpretations, at its core, we believe defect tolerance to mean that the semiconductor maintains free charge-carriers, low non-radiative recombination rates, and high mobilities despite the introduction of defects, which then affords it with the long charge-carrier transport lengths necessary to approach its optical limits in performance. A wide range of models have been proposed for defect tolerance in LHPs. The most popular are the shallow trapping and dielectric screening models, but they have yielded only few materials that may exhibit defect tolerance, and none that have yet matched the performance of LHP PVs. A key issue is that these models do not account for the strength of cation-anion orbital overlap, or the hybridization between dangling bonds. Detailed studies into binary halide PIMs have proven valuable in refining these design criteria. At the same time, there are still several outstanding questions that need to be addressed.

Firstly, the soft and dynamic nature of the LHP structure has complicated the interpretation of the role of traps via the Shockley-Read-Hall model, questioning whether the trap energy is a reasonable proxy for the carrier capture rate. In-depth investigations combining high-level theory with state-of-the-art characterization that can probe similar length- and time-scales will be necessary to resolve this debate.

Secondly, a greater mechanistic understanding of self-healing is required, which will establish the importance of ion transport both in the dark and under illumination. Currently, self-healing is poorly understood computationally due to the long timescales involved, but there is growing experimental evidence of its importance for the manifestation of defect tolerance in LHPs. If self-healing requires significant mass transport, this would open up important questions regarding whether there would be a compromise with the stability of devices under operation.

Thirdly, it will be critical to consider the role of carrier-phonon coupling. This includes understanding whether a trade-off is required between factors affecting both defect tolerance and carrier localization. For example, lower-energy phonon modes contribute to defect tolerance, but would increase the strength of Fröhlich coupling, reducing mobilities. Similarly, higher Born-effective charges increase the ionic contribution to the dielectric constant, which leads to improved screening of carriers from charged defects, but would also increase Fröhlich coupling. Beyond the current focus on band-edge charge-carriers, an important challenge will be to extend the understanding of defect tolerance to hot carriers. In particular, understanding how traps influence the hot carrier cooling process, hot phonon bottleneck effect, Auger reheating and the overlap of polarons will be critical to achieving hot carrier solar cells that can overcome the radiative limit.

Overall, addressing these challenges could improve our understanding of the multifaceted nature of defect tolerance, not only for the current focus on intrinsic defects, but also extended to extrinsic impurities[235]. Together, these can lead to new avenues to create efficient and cost-effective semiconductors.

## Acknowledgements

Y.-T.H. and H.L. arranged alphabetically by surname in byline. I. M.-L. acknowledges Imperial College London for funding from a President's PhD scholarship. R. L. Z. H., H. L. and J. Y. acknowledge support from a UK Research and Innovation Frontier Grant (grant no.

EP/X029900/1), awarded through the European Research Council Starting Grant 2021 scheme. H. L. thanks the Department of Chemistry at the University of Oxford for a studentship. R. L. Z. H. and Y.-T. H. thank the Engineering and Physical Sciences Research Council (EPSRC, grant no. EP/V014498/2) for financial support. A. W. is supported by EPSRC project no. EP/X037754/1. R. L. Z. H. thanks the Royal Academy of Engineering through the Senior Research Fellowships scheme for financial support (grant no. RCSRF2324-18-68).

## Author contributions

R. L. Z. H. and A. W. conceived of the idea for this review and drafted the proposal, with support from the other authors. R. L. Z. H. wrote the introduction, Box 1 and Box 3, the definition of defect tolerance section, drafted the conclusions and outlook, and contributed to Fig. 1. I.M.-L. and A.W. wrote the models for defect tolerance in LHPs, prepared Box 2, and contributed to Fig. 1. J.Y. wrote the polaronic model sub-section in the main discussion and outlook section. H.L. prepared Fig. 3, and the discussion around that, while Y.-T.H. prepared Fig. 4 and the associated discussion. All authors edited and revised the manuscript together.

## Competing interests

The authors declare no competing interests

**Supplementary Information for**

# Multifaceted nature of defect tolerance in halide perovskites and emerging semiconductors

Irea Mosquera-Lois[1,†], Yi-Teng Huang[2,†], Hugh Lohan[2,1,†], Junzhi Ye[2], Aron Walsh[1,*] and Robert L. Z. Hoye[2,*]

[1.] Department of Materials, Imperial College London, Exhibition Road, London SW1 2AZ, United Kingdom

[2.] Inorganic Chemistry Laboratory, Department of Chemistry, University of Oxford, South Parks Road, Oxford OX1 3QR, United Kingdom

[†] These authors contributed equally

* e-mail: a.walsh@imperial.ac.uk (A.W.), robert.hoye@chem.ox.ac.uk (R. L. Z. H.)

**Table S1 | Performance of lead-free perovskite-inspired materials in solar cells** (1-sun illumination)

| Year | Solar absorber | Device architecture | $V_{OC}$ (V) | $J_{SC}$ (mA cm$^{-2}$) | FF (%) | PCE (%) | Ref. |
|---|---|---|---|---|---|---|---|
| 2015 | $FASnI_3$ | FTO/c-$TiO_2$/mp-$TiO_2$/$FASnI_3$/spiro-OMeTAD/Au | 0.238 | 24.45 | 36 | 2.1 | 1 |
| 2016 | $FASnI_3$ | FTO/c-$TiO_2$/mp-TiO2/$FASnI_3$/Spiro-OMeTAD/Au | 0.32 | 23.7 | 63 | 4.8 | 2 |
| 2016 | $FASnI_3$ | FTO/c-$TiO_2$/mp-$TiO_2$/ZnS/$FASnI_3$/PTAA/Au | 0.38 | 23.09 | 60.01 | 5.27 | 3 |
| 2017 | $FASnI_3$ | FTO/c-$TiO_2$/mp-TiO2/$FASnI_3$/PTAA/Au | 0.48 | 22.54 | 64.47 | 7.14 | 4 |
| 2018 | $FASnI_3$ | ITO/PEDOT:PSS/$FASnI_3$/BCP/Al | 0.525 | 24.1 | 0.71 | 9 | 5 |
| 2019 | $FASnI_3$ | FTO/NiO/$FASnI_3$/$C_{60}$/Au | 0.97 | 25.95 | 80.85 | 9.99 | 6 |
| 2020 | $FASnI_3$ | ITO/PEDOT:PSS/$FASnI_3$/ $C_{60}$/BCP/Ag | 0.638 | 21.95 | 72.5 | 10.16 | 7 |
| 2020 | $FASnI_3$ | ITO/PEDOT:PSS/$FASnI_3$/PMMA/ BCP/Ag | 0.67 | 22.47 | 71.8 | 10.81 | 8 |
| 2020 | $FASnI_3$ | ITO/PEDOT:PSS/$FASnI_3$/ $C_{60}$/BCP/Ag | 0.76 | 23.5 | 64 | 11.4 | 9 |
| 2021 | $FASnI_3$ | ITO/PEDOT:PSS/$FASnI_3$/ $C_{60}$/BCP/Ag | 0.77 | 22.48 | 70 | 12.11 | 10 |
| 2021 | $FASnI_3$ | ITO/PEDOT:PSS/$FASnI_3$/ICBA/ BCP/Al | 0.84 | 24.91 | 71 | 14.81 | 11 |
| 2024 | $FASnI_3$ | ITO/PEDOT/$FASnI_3$/BCP/Ag | 0.974 | 21.7 | 74.4 | 15.7 | 12 |
| 2017 | $MASnI_3$ | FTO/c-$TiO_2$/mp-$TiO_2$/{en}$MASnI_3$:$SnF_2$/PTAA/Au | 0.428 67 | 24.28 | 63.72 | 6.63 | 13 |
| 2019 | $MASnI_3$ | FTO/c-$TiO_2$/mp-$TiO_2$/$MASnI_3$/PTAA/Au | 0.426 | 22.91 | 64 | 7.13 | 14 |
| 2020 | $MASnI_3$ | ITO/PEDOT:PSS/$MASnI_3$/$PC_{61}BM$/ BCP/Ag | 0.57 | 20.68 | 66 | 7.78 | 15 |
| 2022 | $MASnI_3$ | ITO/PEDOT:PSS/$MASnI_3$-*x*EABr/$PC_{61}BM$/BCP/Ag | 0.72 | 19.08 | 69.62 | 9.59 | 16 |
| 2024 | $MASnI_3$ | ITO/PEDOT:PSS/$MASnI_3$/$PC_{61}BM$/ BCP/Ag | 0.74 | 22.3 | 62 | 10.2 | 17 |
| 2016 | $CsSnI_3$ | ITO/$TiO_2$/m-$TiO_2$/$CsSnI_3$/Spiro-OMeTAD/Au | 0.52 | 10.2 | 62.5 | 3.31 | 18 |
| 2016 | $CsSnI_3$ | ITO/$CsSnI_3$/$PC_{61}BM$/BCP/Al | 0.51 | 10.44 | 69 | 3.56 | 19 |
| 2017 | $CsSnI_3$ | FTO/c-$TiO_2$/mp-$TiO_2$/$CsSnI_3$/PTAA:TPFB/Au | 0.381 | 25.71 | 49.05 | 4.81 | 20 |
| 2019 | $CsSnI_3$ | ITO/PEDOT:PSS/$CsSnI_3$/ PCBM/Ag | - | - | - | 5.03 | 21 |
| 2021 | $CsSnI_3$ | FTO/c-TiO2/mp-$TiO_2$/$CsSnI_3$-MBAA/P3HT/Au | 0.45 | 24.85 | 67 | 7.5 | 22 |
| 2021 | $CsSnI_3$ | ITO/PEDOT:PSS/$CsSnI_3$-PTM/ICBA/BCP/Ag | 0.64 | 21.81 | 72.1 | 10.1 | 23 |
| 2022 | $CsSnI_3$ | ITO/$NiO_x$/$CsSnI_3$/PCBM/ ZrAcac/Ag | 0.75 | 20.7 | 72.1 | 11.2 | 24 |
| 2023 | $CsSnI_3$ | Ag/Spiro-OMeTAD/$CsSnI_3$/PCBM/Ag | 0.78 | 25.13 | 59.8 | 11.7 | 25 |
| 2024 | $CsSnI_3$ | ITO/PEDOT:PSS/$CsSnI_3$/ 3-ThMAI/ICBA/BCP/Ag | 0.77 | 22.68 | 69 | 12.05 | 26 |
| 2016 | $MA_3Bi_2I_9$ | ITO/PEDOT/$MA_3Bi_2I_9$/PCBM/Ca/Al | 0.66 | 0.22 | 49 | 0.1 | 27 |
| 2016 | $MA_3Bi_2I_9$ | ITO/c-$TiO_2$/mp$TiO_2$/$MA_3Bi_2I_9$/Spiro-MeOTAD/$MoO_3$/Ag | 0.67 | 1 | 62.48 | 0.42 | 28 |
| 2018 | $MA_3Bi_2I_9$ | FTO/c-$TiO_2$/mp-$TiO_2$/$MA_3Bi_2I_9$/P3HT/Au | 1.01 | 4.02 | 78 | 3.17 | 29 |
| 2015 | $Cs_3Bi_2I_9$ | FTO/c-$TiO_2$/mp-$TiO_2$/$Cs_3Bi_2I_9$/ Spiro-OMeTAD/Ag | 0.85 | 2.15 | 60 | 1.09 | 30 |
| 2016 | $Cs_3Bi_2I_9$ | FTO/c-$TiO_2$/mp-$TiO_2$/$Cs_3Bi_2I_9$/P3HT/Ag | 0.26 | 0.18 | 37 | 0.02 | 31 |
| 2018 | $Cs_3Bi_2I_9$ | FTO/mp-$TiO_2$/$Cs_3Bi_2I_9$/Spiro-OMeTAD/Au | 0.49 | 0.67 | 63.6 | 0.21 | 32 |
| 2018 | $Cs_3Bi_2I_9$ | FTO/$TiO_2$/$Cs_3Bi_2I_9$/CuI/Au | 0.86 | 5.78 | 64.38 | 3.20 | 33 |
| 2019 | $Cs_3Bi_2I_9$ | ITO/$NiO_x$/$Cs_3Bi_2I_9$/PCBM/ Al:ZnO/Ag | 0.74 | 3.42 | 51 | 1.26 | 34 |
| 2019 | $Cs_3Bi_2I_9$ | ITO/$NiO_x$/$Cs_3Bi_2I_9$/PCBM/$C_{60}$/ BCP/Ag | 0.75 | 0.51 | 59 | 0.23 | 35 |
| 2019 | $Cs_3Bi_2I_{9-x}Br_x$ | ITO/$NiO_x$/$Cs_3Bi_2I_9$/PCBM/$C_{60}$/ BCP/Ag | 0.64 | 3.15 | 57 | 1.15 | 35 |
| 2020 | $Cs_3Bi_2I_9$ | Al:ZnO/c-$TiO_2$/$Cs_3Bi_2I_9$/CuSCN/graphite | 0.37 | 1.43 | 32 | 0.17 | 36 |
| 2020 | $Cs_3Bi_2I_9$ | FTO/c-$TiO_2$/mp-$TiO_2$/$Cs_3Bi_2I_9$/PDBD-T/Au | 0.86 | 0.25 | 35.27 | 0.08 | 37 |
| 2020 | $Cs_3Bi_2I_9$ | FTO/$TiO_2$/$Cs_3Bi_2I_9$/CuI/Au | 0.675 | 5.39 | 52 | 1.087 | 38 |
| 2022 | $Cs_3Bi_2I_9$ | FTO/c-$TiO_2$/mp-$TiO_2$/$Cs_3Bi_2I_9$/C | 1.01 | 3.60 | 77 | 2.81 | 39 |
| 2022 | $Cs_3Bi_2I_9$ | FTO/c-$TiO_2$/mp-$TiO_2$/$Cs_3Bi_2I_9$/Spiro-OMeTAD/Au | 0.50 | 3.58 | 54.42 | 0.98 | 40 |
| 2023 | $Cs_3Bi_2I_9$ | FTO/$TiO_2$/$Cs_3Bi_2I_9$/CuI/Au | 1.07 | 2.28 | 62.30 | 1.52 | 41 |
| 2022 | $Cs_2AgBiBr_6$ | ITO/$SnO_2$/$Cs_2AgBiBr_6$/Spiro-OMeTAD/Au | 0.92 | 11.4 | 60.93 | 6.37 | 42 |
| 2018 | $Cs_3Sb_2I_9$ | FTO/c-$TiO_2$/mp-$TiO_2$/$Cs_3Sb_2I_9$/Spiro-OMeTAD/Au | 0.404 | 0.13 | 58 | 0.03 | 43 |
| 2018 | $Cs_3Sb_2I_9$ | ITO/PEDOT:PSS/ $Cs_3Sb_2I_9$/$PC_{70}BM$/Al | 0.72 | 5.31 | 38.97 | 1.49 | 44 |
| 2019 | $Cs_3Sb_2I_9$ | FTO/c-$TiO_2$/$Cs_3Sb_2I_9$/Au | 0.61 | 3.55 | 55.8 | 1.21 | 45 |
| 2020 | $Cs_3Sb_2I_9$ | ITO/PEDOT:PSS/ $Cs_3Sb_2I_9$/$PC_{71}BM$/Al | 0.79 | 3.76 | 54 | 1.67 | 46 |
| 2020 | $Cs_3Sb_2I_9$ | FTO/$TiO_2$/$Cs_3Sb_2I_9$/CuI/Au | 0.653 | 4.56 | 46 | 1.036 | 47 |
| 2020 | $Cs_3Sb_2I_{9-x}Cl_x$ | ITO/PEDOT:PSS/$Cs_3Sb_2I_{9-x}Cl_x$ /$PC_{60}BM$/Al | 0.70 | 5.87 | 42 | 1.7 | 48 |
| 2021 | $Cs_3Sb_2I_9$ | ITO/PEDOT:PSS/ $Cs_3Sb_2I_9$/ITIC/PCBM/Ca/Ag | 0.91 | 6.45 | 55 | 3.25 | 49 |
| 2022 | $Cs_3Sb_2I_9$ | FTO/$TiO_2$/$Cs_3Sb_2I_9$/P3HT/Au | 0.80 | 5.40 | 54.90 | 2.48 | 50 |
| 2022 | $Cs_3Sb_2I_9$ | FTO/$TiO_2$/$Cs_3Sb_2I_9$/Spiro-OMeTAD/Au | 0.622 | 3.69 | 47 | 1.07 | 51 |
| 2022 | $Cs_3Sb_2I_{9-x}Cl_x$ | FTO/$Nb_2O_5$/$Cs_3Sb_2I_{9-x}Cl_x$ /P3HT/C | 0.80 | 3.87 | 53 | 1.67 | 52 |
| 2022 | $Cs_3Sb_2I_{9-x}Cl_x$ | FTO/c-$TiO_2$/mp-$TiO_2$/$Cs_3Sb_2I_{9-x}Cl_x$ /Spiro-OMeTAD/Au | 0.65 | 6.77 | 50.3 | 2.22 | 53 |
| 2022 | $Cs_3Sb_2I_{9-x}Cl_x$ | ITO/PEDOT:PSS/$Cs_3Sb_2I_{9-x}Cl_x$ /PMMA/PCBM/Al | 0.87 | 6.55 | 50 | 2.85 | 54 |
| 2023 | $Cs_3Sb_2I_9$ | FTO/c-$TiO_2$/$Cs_3Sb_2I_9$/Au | 0.62 | 5.57 | 51.4 | 1.76 | 55 |
| 2023 | $Cs_3Sb_2I_{9-x}Cl_x$ | FTO/a-$Nb_2O_5$/ $Cs_3Sb_2I_{9-x}Cl_x$/P3HT/C | 0.83 | 4.04 | 52 | 1.75 | 56 |
| 2023 | $Cs_3Sb_2I_{9-x}Cl_x$ | FTO/$TiO_2$/$Cs_3Sb_2I_{9-x}Cl_x$ /Spiro-OMeTAD/Au | 0.84 | 7.13 | 54.2 | 3.2 | 57 |
| 2013 | $AgBiS_2$ | FTO/c-$TiO_2$/mp-$TiO_2$/ $AgBiS_2$/($Na_2S$+S+KCl+NaOH)/Au | 0.18 | 7.61 | 38.6 | 0.53 | 58 |
| 2015 | $AgBiS_2$ | FTO/$TiO_2$ NRs/ $AgBiS_2$/(LiI+$I_2$+DMPII+4-TPB)/Pt-coated FTO | 0.53 | 4.22 | 43 | 0.95 | 59 |
| 2016 | $AgBiS_2$ | ITO/ZnO/$AgBiS_2$/PTB7/$MoO_3$/Ag | 0.45 | 22.1 | 65 | 6.3 | 60 |
| 2018 | $AgBiS_2$ | ITO/ZnO/$AgBiS_2$/P3HT/Au | 0.46 | 16.7 | 56 | 4.3 | 61 |
| 2018 | $AgBiS_2$ | ITO/ZnO/$AgBiS_2$/Spiro-OMeTAD/$MoO_3$/Ag | 0.241 | 18.1 | 35 | 1.5 | 62 |
| 2019 | $AgBiS_2$ | FTO/TiO2/$AgBiS_2$/Co:P3HT/C/Ag | 0.50 | 13.27 | 43 | 2.87 | 63 |
| 2019 | $AgBiS_2$ | FTO/ZnO/$AgBiS_2$/P3HT/$MoO_3$/Al | 0.23 | 16.56 | 36.75 | 1.40 | 64 |
| 2019 | $AgBiS_2$ | ITO/ZnO/$AgBiS_2$/PTB7/$MoO_3$/Ag | 0.51 | 17.63 | 64 | 5.75 | 65 |
| 2020 | $AgBiS_2$ | ITO/ZnO/$AgBiS_2$/PTB7/$MoO_3$/Ag | 0.46 | 22.68 | 61 | 6.37 | 66 |
| 2020 | $AgBiS_2$ | ITO/ZnO/$AgBiS_2$/P3HT/$MoO_3$/Au | 0.41 | 15.06 | 54 | 3.31 | 67 |
| 2020 | $AgBiS_2$ | ITO/PEDOT:PSS/$AgBiS_2$/BCP/$C_{60}$/ Cu | 0.28 | 16.85 | 43.3 | 2.04 | 68 |
| 2020 | $AgBiS_2$ | FTO/ZnO/$AgBiS_2$/ PTB7-Th/$MoO_3$/Ag | 0.445 | 18.87 | 54.4 | 4.57 | 69 |
| 2020 | $AgBiS_2$ | ITO/ZnO/$AgBiS_2$/PTB7/$MoO_x$/Ag | 0.43 | 22.07 | 59 | 5.55 | 70 |
| 2020 | $AgBiS_2$ | ITO/ZnO/$AgBiS_2$/PTB7/$MoO_x$/Ag | 0.55 | 12.41 | 59 | 4.08 | 71 |
| 2021 | $AgBiS_2$ | ITO/ZnO NWs/$AgBiS_2$/P3HT/Au | 0.378 | 20.54 | 55.3 | 4.14 | 72 |
| 2021 | $AgBiS_2$ | ITO/NiO/$AgBiS_2$/ZnO/Al | 0.50 | 18.54 | 61 | 5.59 | 73 |
| 2021 | $AgBiS_2$ | ITO/NiO/$AgBiS_2$/PCBM/BCP/Ag | 0.38 | 20.71 | 54 | 4.25 | 74 |
| 2022 | $AgBiS_2$ | FTO/c-$TiO_2$/mp-$TiO_2$/ $AgBiS_2$/Spiro-OMeTAD:P3HT | 0.35 | 4.59 | 56.47 | 0.91 | 75 |
| 2022 | $AgBiS_2$ | ITO/ZnO/$AgBiS_2$/P3HT/Au | 0.436 | 12.64 | 56.7 | 3.12 | 76 |
| 2022 | $AgBiS_2$ | ITO/ZnO/$AgBiS_2$/ PBDB-T-2F/$MoO_3$/Ag | 0.494 | 27.07 | 68.1 | 9.10 | 77 |
| 2022 | $AgBiS_2$ | ITO/$SnO_2$/$AgBiS_2$/PTAA/$MoO_3$/Ag | 0.495 | 27.11 | 68 | 9.17 | 78 |
| 2022 | $AgBiS_2$ | FTO/TiO2/$AgBiS_2$/NiO/CuS | 0.24 | 30.2 | 29.8 | 2.1 | 79 |
| 2022 | $AgBiS_2$ | ITO/$SnO_2$/$AgBiS_2$/PTAA/$MoO_3$/Ag | 0.48 | 24.9 | 61 | 7.3 | 80 |
| 2022 | $AgBiS_2$ | ITO/ZnO-NWs/$AgBiS_2$/P3HT/Au | 0.41 | 22.21 | 60 | 5.41 | 81 |
| 2022 | $AgBiS_2$ | ITO/$SnO_2$/$AgBiS_2$/PTB7/$MoO_3$/Ag | 0.48 | 23.97 | 64 | 7.33 | 82 |
| 2022 | $AgBiS_2$ | FTO/TiON/$AgBiS_2$/NiO/CuS | 0.30 | 16.1 | 49.7 | 2.0 | 83 |
| 2023 | $AgBiS_2$ | FTO/c-$TiO_2$/mp-$TiO_2$/$AgBiS_2$/ (N719+$I_2$)/Pt | 0.78 | 14.87 | 72 | 8.36 | 84 |
| 2024 | $AgBiS_2$ | SLG/Mo/$AgBiS_2$/CdS/ITO/Ag | 0.096 5 | 4.68 | 26 | 0.13 | 85 |
| 2024 | $AgBiS_2$ | FTO/$SnO_2$/$AgBiS_2$/Spiro- OMeTAD/Au | 0.28 | 30.9 | 41.9 | 3.68 | 86 |
| 2024 | $AgBiS_2$ | ITO/$SnO_2$/$AgBiS_2$/PTAA/$MoO_3$/Ag | 0.486 | 23.81 | 64 | 7.35 | 87 |
| 2024 | $AgBiS_2$ | ITO/ZnO/$AgBiS_2$/PTB7/$MoO_x$/Ag | 0.54 | 22.6 | 67.44 | 8.14 | 88 |
| 2024 | $AgBiS_2$ | ITO/$SnO_2$/$AgBiS_2$/PTAA/$MoO_3$/Au | 0.518 | 27.20 | 72.4 | 10.20 | 89 |
| 2020 | $BiI_3$ | ITO/PEDOT/$BiI_3$+PCBM/BCP/Ag | 0.47 | 8.76 | 36.5 | 1.50 | 90 |
| 2010 | BiOI | FTO/$TiO_2$/BiOI/electrolyte/Pt/FTO | 0.62 | 0.24 | 61 | 0.092 | 91 |
| 2015 | BiOI | FTO/$TiO_2$/BiOI/electrolyte/Pt/FTO | 0.61 | 3.8 | 45 | 1.03 | 92 |
| 2017 | BiOI | ITO/$NiO_x$/BiOI/ZnO/Al | 0.75 | 7.0 | 43 | 1.82 | 93 |
| 2019 | SbSI | FTO/c-$TiO_2$/mp- $TiO_2$/SbSI/PCPDTBT/Au | 0.60 | 9.62 | 65.2 | 3.62 | 94 |
| 2014 | $CuSbS_2$ | Mo/$CuSbS_2$/CdS/Al:ZnO | 0.49 | 14.73 | 44 | 3.13 | 95 |
| 2016 | $CuSbS_2$ | Mo/$CuSbS_2$/CdS/i-ZnO/n-ZnO/Al | 0.47 | 15.6 | 43.6 | 3.22 | 96 |
| 2015 | $CuSbSe_2$ | FTO/$CuSbSe_2$ /CdS/ZnO/ITO/Al | 0.274 | 11.84 | 40.51 | 1.32 | 97 |
| 2015 | $CuSbSe_2$ | Mo/$CuSbSe_2$/CdS/i- ZnO/ZnO/Ni:Al/MgF | 0.35 | 22.82 | - | 3.5 | 98 |
| 2017 | $CuSbSe_2$ | Mo/$CuSbSe_2$/CdS/i- ZnO/ZnO/Ni:Al/MgF | 0.336 | 26 | 53 | 4.7 | 99 |
| 2014 | $Sb_2Se_3$ | FTO/c-$TiO_2$/m-$TiO_2$/$Sb_2Se_3$ /PEDOT:PSS/Au | 0.304 5 | 22.3 | 47.2 | 3.21 | 100 |
| 2014 | $Sb_2Se_3$ | FTO/CdS/$Sb_2Se_3$/Au | 0.335 | 24.4 | 46.8 | 3.7 | 101 |
| 2015 | $Sb_2Se_3$ | ITO/CdS/$Sb_2Se_3$/Au | 0.36 | 25.3 | 52.5 | 4.8 | 102 |
| 2015 | $Sb_2Se_3$ | FTO/CdS/$Sb_2Se_3$/Au | 0.4 | 25.1 | 55.7 | 5.6 | 103 |
| 2017 | $Sb_2Se_3$ | $TiO_2$/CdS/GQDs–C-fabric/NiO/ $Sb_2Se_3$ | 0.63 | 25.49 | 45 | 7.19 | 104 |
| 2019 | $Sb_2Se_3$ | ZnO:Al/ZnO/CdS/TiO2/$Sb_2Se_3$- NRs/MoSe2/Mo | 0.4 | 32.58 | 70.3 | 9.2 | 105 |
| 2022 | $Sb_2Se_3$ | Mo/$MoSe_2$/ $Sb_2Se_3$/CdS/ZnO/AZO | 0.488 | 30.86 | 67.17 | 10.12 | 106 |
| 2022 | $Sb_2Se_3$ | FTO/CdS/$Sb_2Se_3$/spiro- OMeTAD/Au | 0.767 | 33.52 | 67.64 | 10.57 | 107 |
| 2009 | $Sb_2S_3$ | FTO/TiO2/$Sb_2Se_3$/CuSCN/Au | 0.49 | 14.1 | 48.8 | 3.37 | 108 |
| 2010 | $Sb_2S_3$ | FTO/$TiO_2$/ $Sb_2S_3$/(LiSCN)CuSCN/Au | 0.56 | 11.6 | 58 | 3.7 | 109 |
| 2010 | $Sb_2S_3$ | FTO/$TiO_2$/$Sb_2S_3$/Spiro/Au | 0.556 | 12.3 | 69.9 | 5.06 | 110 |
| 2011 | $Sb_2S_3$ | FTO/$TiO_2$/ $Sb_2S_3$/PCDTBT/PTAA/Au | 0.616 | 15.3 | 65.7 | 6.18 | 111 |
| 2012 | $Sb_2S_3$ | FTO/$TiO_2$/$Sb_2S_3$/P3HT/PCBM/Au | 0.595 | 16 | 65.5 | 6.3 | 112 |
| 2014 | $Sb_2S_3$ | FTO/mp- $TiO_2$/$Sb_2S_3$/PCPDTBT/PEDOT:PS S/Au | 0.711 | 16.1 | 65 | 7.5 | 113 |
| 2015 | $Sb_2S_3$ | FTO/$TiO_2$/$Sb_2S_3$/P3HT/PEDOT:PS S/Au | 0.595 | 16.1 | 66.5 | 6.4 | 114 |
| 2018 | $Sb_2S_3$ | FTO/TiO2/$Sb_2S_3$/Spiro- OMeTAD/Au | 0.671 | 18.43 | 54.8 | 6.78 | 115 |
| 2020 | $Sb_2S_3$ | FTO/$TiO_2$/ $Sb_2S_3$/PCPDTBT/PEDOT:PSS/Au | 0.521 | 21.5 | 63 | 7.05 | 116 |
| 2020 | $Sb_2S_3$ | FTO/$TiO_2$/$SbCl_3$-treated $Sb_2S_3$/Spiro-OMeTAD/Au | 0.72 | 17.24 | 57.18 | 7.1 | 117 |
| 2022 | $Sb_2S_3$ | FTO/CdS/$Sb_2S_3$/Spiro-OMeTAD/Au | 0.757 | 17.41 | 60.48 | 8 | 118 |
| 2023 | $Sb_2S_3$ | FTO/$SnO_2$/CdS/$Sb_2S_3$/Spiro-OMeTAD/Au | 0.796 | 16.7 | 57.7 | 7.66 | 119 |
| 2023 | $Sb_2S_3$ | ITO/$TiO_2$/CdS/$Sb_2S_3$/ Spiro-OMeTAD/$MoO_3$/Ag | 0.75 | 17.18 | 60 | 7.73 | 120 |
| 2009 | $Sb_2(S_xSe_{1-x})_3$ | d-$SnO_2$/CdS/ $Sb_2(S_xSe_{1-x})_3$:$Sb_2O_3$/PbS | 0.52 | 4.2 | 28 | 0.66 | 121 |
| 2014 | $Sb_2(S_xSe_{1-x})_3$ | FTO/c-$TiO_2$/mp-$TiO_2$/$Sb_2S_3$/$Sb_2Se_3$/P3HT/Au | 0.474 8 | 24.9 | 55.6 | 6.6 | 122 |
| 2019 | $Sb_2(S_xSe_{1-x})_3$ | FTO/$TiO_2$/CdS:In/ $Sb_2(S_xSe_{1-x})_3$/Spiro-OMeTAD/Au | 0.59 | 18.1 | 62 | 6.63 | 123 |
| 2020 | $Sb_2(S_xSe_{1-x})_3$ | FTO/CdS/$Sb_2(S_xSe_{1-x})_3$/Spiro-OMeTAD/ Au | 0.63 | 23.7 | 68 | 10 | 124 |
| 2020 | $Sb_2(S_xSe_{1-x})_3$ | FTO/CdS/$Sb_2(S_xSe_{1-x})_3$/Spiro-OMeTAD/Au | 0.664 | 23.8 | 66 | 10.5 | 125 |
| 2022 | $Sb_2(S_xSe_{1-x})_3$ | FTO/Zn(O,S)/CdS/ $Sb_2(S_xSe_{1-x})_3$/Spiro-OMeTAD/Au | 0.673 | 23.7 | 67 | 10.7 | 126 |
| 2023 | $Sb_2(S_xSe_{1-x})_3$ | glass/FTO/CdS/ $Sb_2(S_xSe_{1-x})_3$/Spiro-OMeTAD/Au | 0.63 | 25.27 | 67.35 | 10.75 | 127 |

**Abbreviations** - mesoporous $TiO_2$ (mp-$TiO_2$), compact $TiO_2$ (c-$TiO_2$),Graphene Quantum Dots (GQDs), $Al_2O_3$:ZnO (AZO), Nanorods (NRs), doped $SnO_2$ (d-$SnO_2$), phthalimide (PTM), 3-thiophenemethylammonium iodide (3-ThMAI), indene-$C_{60}$ bisadduct (ICBA), bahocuproine (BCP), (Spiro-OMeTAD), Poly[2,6-(4,4-bis-(2-ethylhexyl)-4*H*-cyclopenta[2,1-b;3,4-b']dithiophene)-*alt*-4,7(2,1,3-benzothiadiazole)] (PCPDTBT), Poly(3,4-ethylenedioxythiophene) polystyrene sulfonate (PEDOT:PSS), Poly[bis(4-phenyl)(2,4,6-trimethylphenyl)amine (PTAA), Poly({4,8-bis[(2-ethylhexyl)oxy]benzo[1,2-b:4,5-b′]dithiophene-2,6-diyl}{3-fluoro-2-[(2-ethylhexyl)carbonyl]thieno[3,4-b]thiophenediyl}) (PTB7), [6,6]-Phenyl C61 butyric acid methyl ester (PCBM), Poly[(2,6-(4,8-bis(5-(2-ethylhexyl)thiophen-2-yl)-benzo[1,2-b:4,5-b']dithiophene))-alt-(5,5-(1',3'-di-2-thienyl-5',7'-bis(2-ethylhexyl)benzo[1',2'-c:4',5'-c']dithiophene-4,8-dione)] (PDBD-T), N,N'-methylenebis(acrylamide) (MBAA), ethylammonium bromide (EABr), Poly[[4,8-bis[5-(2-ethylhexyl)-4-fluoro-2-thienyl]benzo[1,2-b:4,5-b′ ]dithiophene-2,6-diyl]-2,5-thiophenediyl[5,7-bis(2-ethylhexyl)-4,8-dioxo-4H,8H-benzo[1,2-c:4,5-c′ ]dithiophene-1,3-diyl]-2,5-thiophenediyl] (PBDB-T-2F)